\documentclass{osa-article}

\journal{oe}

\usepackage{float}
\articletype{Research Article}

\begin{document}

\title{MetaNet: A new paradigm for data sharing in photonics research}

\author{Jiaqi Jiang,\authormark{1} Robert Lupoiu,\authormark{2} Evan W. Wang,\authormark{1} David Sell,\authormark{3} Jean Paul Hugonin,\authormark{4} Philippe Lalanne,\authormark{5} and Jonathan A. Fan\authormark{1}*}

\address{\authormark{1}Department of Electrical Engineering, Stanford University, Stanford, CA 94305, USA\\
\authormark{2}Department of Electrical Engineering, University of British Columbia, Vancouver, BC V6T 1Z4, Canada\\
\authormark{3}Department of Applied Physics, Stanford University, Stanford, CA 94305, USA\\
\authormark{4}Laboratoire Charles Fabry,  Institut d'Optique Graduate School, Universit\'{e} Paris-Saclay, Paris, France\\
\authormark{5}Laboratoire Photonique,  Institut d'Optique Graduate School, Universit\'{e} Bordeaux, Bordeaux, France}

\email{\authormark{*}jonfan@stanford.edu} 



\begin{abstract}
Optimization methods are playing an increasingly important role in all facets of photonics engineering, from integrated photonics to free space diffractive optics.  However, efforts in the photonics community to develop optimization algorithms remain uncoordinated, which has hindered proper benchmarking of design approaches and access to device designs based on optimization.  We introduce MetaNet, an online database of photonic devices and design codes intended to promote coordination and collaboration within the photonics community.  Using metagratings as a model system, we have uploaded over one hundred thousand device layouts to the database, as well as source code for implementations of local and global topology optimization methods.  Further analyses of these large datasets allow the distribution of optimized devices to be visualized for a given optimization method.  We expect that the coordinated research efforts enabled by MetaNet will expedite algorithm development for photonics design.
\end{abstract}

\section{Introduction}

Photonics engineering, in which materials are geometrically structured to tailor their near- and far-field electromagnetic responses, has greatly shaped many technological domains, including communications \cite{vlasov2012silicon, haffner2015all, pospischil2013cmos, yang2017low}, image sensors\cite{wang2019single, tittl2018imaging, chen2012metamaterials, singh2014ultrasensitive, lee2002imaging}, energy harvesting \cite{weintraub2009optical, park2011photonic, koenderink2015nanophotonics, baranov2019nanophotonic}, and medical diagnostics \cite{conde2012nanophotonics, rosa2012gold}.  Traditionally, photonics engineering has been driven by design concepts based on relatively simple geometries.  For example, elements in planar chip-based photonic systems, such as ring resonators, mode converters, and demultiplexers, consist of simple shapes based on the coupling or tapering of waveguide structures.  Similarly, diffractive optics in the form of metasurfaces \cite{yu2014flat, yang2018freeform, kildishev2013planar, phan2019high, henstridge2018accelerating} and metamaterials utilize `meta-atoms,' which consist of nanoresonators or nanowaveguides with intuitive optical responses.  The use of simple shapes is advantageous as it enables the design process to be guided by physical intuition and allows devices to be realized using relatively simple design rules.  Importantly, these design rules are typically captured through simple mathematical expressions, which allows these concepts to be easily disseminated and reproduced.

Traditional device designs are an effective solution for many problems and will continue to drive much photonics technology development.  However, they encompass only a small portion of the total design space, and it is clear that this restriction leads to performance limitations for many device types.  To overcome these limits, researchers have been actively developing optimization algorithms to more thoroughly explore the full design space.  These algorithms range from conventional optimization methods, such as evolutionary, annealing\cite{zhao2016broadband}, and genetic algorithms \cite{shen2003design}, to those based on gradient-based topology optimization \cite{sell2018ultra, sell2017large, wang2019robust}.  More recently, a number of groups have harnessed machine learning as a means to facilitate electromagnetic simulations and the design process\cite{jiang2019free, peurifoy2018nanophotonic, inampudi2018neural, liu2018training, liu2018generative, hughes2019wave}.

These collective efforts have given rise to new classes of high-performance photonic design technologies based on freeform layouts\cite{molesky2018inverse, lalau2013adjoint, hughes2018adjoint, sell2017large}.  However, these research efforts have been uncoordinated, and many algorithms and device layouts remain unshared.  Without the sharing of algorithms and their application to common problems, it has been difficult, if not impossible, for research groups to properly benchmark different optimization methods.  It has also been infeasible for groups to simulate and analyze freeform devices reported by other groups, as these devices can only be properly described using detailed layout files.  Additionally, the lack of coordination within the community prevents the collective production of very large datasets of related devices for applications such as data-driven analysis.

In this paper, we introduce MetaNet (\href{http://metanet.stanford.edu}{http://metanet.stanford.edu}), an online repository of design algorithms and device layouts for photonic technologies.  The inspiration for MetaNet comes from the data sciences community, where swift progress in machine learning development has been largely mediated through a culture of sharing, benchmarking, and use of common datasets, such as those featured in ImageNet \cite{deng2009imagenet} and CIFAR \cite{krizhevsky2009learning}.  With MetaNet, we hope to satisfy four major needs within the photonics community.  The first is providing an accessible and indexed repository for sharing high-performance devices, which can be directly used in an application.  The second is providing labeled large-scale datasets for big data analyses, such as those based on machine learning.  The third is the standardization of conventions and parameters for easy design benchmarking.  The fourth is a centralized resource for openly available nanophotonics design software from groups across the world.  To initiate this effort, we focus here on filling out MetaNet with metagrating devices and codebases, though our goal is to extend MetaNet to many classes of photonic technologies.  Such a culture of sharing and collaboration is currently not mainstream in the photonics community, but we anticipate that this cultural shift will greatly expedite collaboration and the advancement of photonics design tools.

\section{The MetaNet database}

Our focus in this paper is on metasurfaces, specifically periodic metasurfaces (i.e., metagratings) that diffract incident light to the +1 diffraction channel.  Metagratings are an excellent model system for metasurface analysis and algorithm benchmarking for a few reasons.  First, the design objectives are well-defined and unambiguous: we want to maximize the intensity of light in a specific diffraction channel.  Other types of devices may have more ambiguous design objectives.  For example, the performance criteria for metalenses can range from maximizing the field magnitude at the focal point to maximizing the light intensity within the central and first side lobes.  Second, metagratings are periodic structures and can therefore be designed using rigorous coupled wave analysis (RCWA), which is a high-speed electromagnetic solver. Third, metagratings are defined by wavelength-scale unit cells, which have relatively small areas and require modest compute times to simulate.  The expediting of the optimization process with fast solvers and small device areas leads to dramatic reductions in time for algorithm development and the generation of large quantities of data.

\subsection{Metagrating standardization and categorization}
There exist many material systems, design methods, and operating wavelength ranges for metasurfaces, so an important task with MetaNet is to catalog devices that are representative of useful and mainstream design concepts.  We choose polycrystalline silicon as the initial material choice for MetaNet due to its ubiquity in the photonics community and its mature fabrication processing.  Specifically, with its high refractive index, silicon can enable metasurfaces that perform a broad range of tasks while maintaining high efficiencies.  For our design objectives, we consider metagratings that deflect to angles ranging from 40 to 80 degrees, which serves as a relatively challenging design problem.  This range was identified in our prior analysis of metagratings as a regime in which conventional designs perform poorly \cite{sell2018ultra}.  The device thicknesses are set to be less than a half-wavelength thick to ensure that the device features have relatively low aspect ratios and are straightforward to fabricate.

\begin{figure}[H]
    \centering
    \includegraphics[width = 100mm]{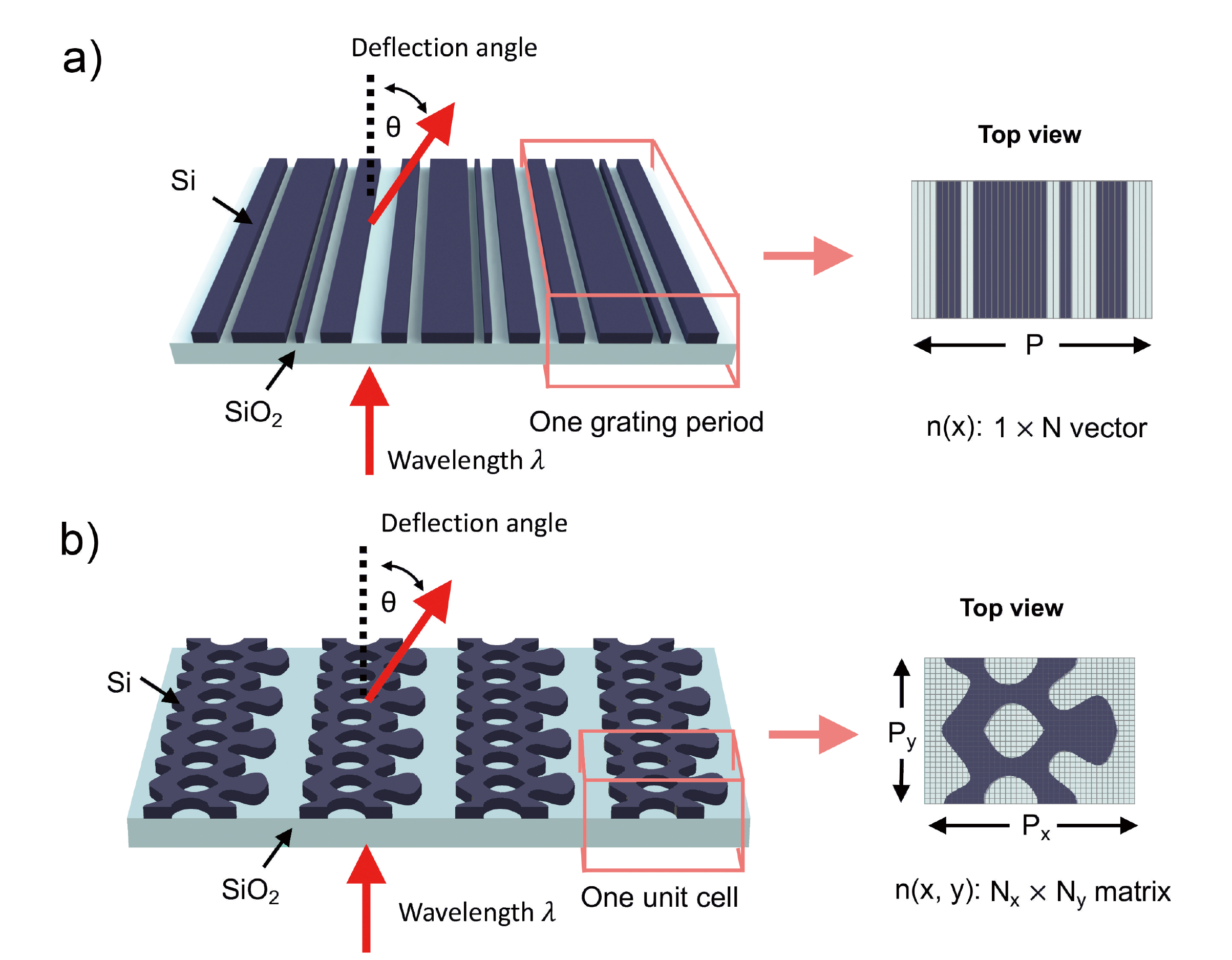}
    \caption{Schematics of (a) 2D and (b) 3D metagrating layouts. The grating unit cells are shown as a top view with an overlayed  grid.}
    \label{fig:layouts}
\end{figure}

The metagratings under consideration can be classified as either 2D or 3D devices.  2D devices consist only of periodic sets of dielectric bars and are desirable due to their relative ease of fabrication, speed of optimization, and robust specification of constraints such as minimum feature size.  Schematics of 2D devices are shown in Figure \ref{fig:layouts}a.  3D devices possess much greater degrees of freedom and utilize free-form structures. This leads to a greatly expanded design space, enhanced performance, and ability to support polarization-independent responses.  Schematics of 3D devices are showing in Figure \ref{fig:layouts}b.

Currently, we have uploaded 135,000 distinct 2D and 3D metagratings that deflect normally incident light to the +1 diffraction order.  Symmetry is imposed such that there is reflection symmetry in the direction normal to the plane of deflection within each metagrating unit cell (Figure 1b).  In the database, devices are labeled and categorized based on operating wavelength, polarization, deflection angle, material, device thickness, dimensionality, design method, and efficiency. It is possible to search and filter devices in MetaNet based on these parameters, and device files can be downloaded individually or as filtered groups. 

For the current batch of uploaded devices, the wavelength range spans 500 to 1300 nm and experimentally-measured refractive indices for polycrystalline silicon are posted in the database.  To simplify our analyses, we have dropped the imaginary part of the refractive index when designing and evaluating our devices.  In the future, we will expand the database to include a wider scope of materials, including other variants of silicon, such as crystalline \cite{2016crystallineSi} and amorphous silicon\cite{arbabi2017planar}, as well as titanium dioxide \cite{ribot2013broadband, devlin2016broadband} and silicon nitride \cite{zhan2016low, yang2017analysis}, which are commonly used for visible light applications.  We will also include devices with more diverse functionalities, such as polarization control, as well as wavelength-scale scattering elements that can be stitched together to create aperiodic structures\cite{phan2019high}.

\subsection{Metagrating optimization algorithms}
We generate metagrating datasets in MetaNet using two different topology optimization methods.  The first is with gradient-based topology optimization based on the adjoint variables method.  This optimization method is local in that it uses the method of gradient descent to improve the performance of an initial dielectric distribution.  The second is with global topology optimization networks (GLOnets), which is a global optimizer.  With GLOnets, the optimization problem is reframed as the training of a generative neural network.  Open source code for both methods is available at: \href{http://metanet.stanford.edu/code/}{http://metanet.stanford.edu/code/}.  We hope that the sharing of baseline optimization code becomes a more mainstream practice within the nanophotonics community, as it would facilitate the advancement of optimization algorithm development.

\paragraph{Gradient-based topology optimization.}  The basis for most of our dataset generation is gradient-based topology optimization.  In this method, the objective is to maximize a Figure of Merit, which in our case is the intensity of light deflected to the desired deflection angle.  For a given device dielectric distribution, the Figure of Merit is iteratively improved by computing a gradient, based on forward and adjoint simulations, and using it to perturb the dielectric values of each voxel in the device each iteration.  For dielectric metagratings, the method is described in more detail in Ref. \cite{sell2018ultra}.  As a method of gradient descent, the optimization itself is deterministic and fully dependent on the initial dielectric distribution.  We generate large datasets by optimizing a batch of devices, each with different random initial dielectric distributions.

In the posted version of this code in MetaNet, we have made efforts to ensure that the code is fast, accurate, readable, and easy to use and modify by researchers unfamiliar with topology optimization.
First, we have organized the code with clear naming conventions. Variable names and definitions are assigned so they are consistent with descriptions of the methods in our prior publications \cite{sell2017large,sell2018ultra} as well as devices currently uploaded to MetaNet.
Second, we include significant in-line documentation to make the code easy to use and modify.  Alongside parameter and function definitions, we have included a set of default hyperparameter values such as the learning rate, binarization coefficient, and robustness-related parameters, so that the code is accessible at all levels of usage and can be run "out of the box."
Third, the code has been structured to be compatible with native Matlab parallel computing capabilities.  With only minor modifications, the code can be run on multiple cores on a single workstation or across multiple nodes in a computing cluster. 

Finally, we couple our algorithms with a mainstream, high-speed, open source electromagnetic solver, Reticolo \cite{lalanne2005reticolo}. Reticolo is Matlab-based and can efficiently and accurately compute the electromagnetic fields required in our optimization.  With our combined optimization and simulation platform, there is no complicated installation or requirement of additional libraries needed to run Reticolo and the entire package can be downloaded from a single location.  The overall setup time from download to running an optimization should take less than five minutes.  In addition to being easily accessible, we have found Reticolo to be amongst the fastest and most accurate RCWA solvers available.  Reticolo is capable of running efficiently on different computing systems ranging from personal computers to computing clusters, without the need for expensive GPUs and excessive memory, which is an often overlooked feature in newer electromagnetic solvers.

At present, our code can be configured to apply to a broad range of metagrating optimization cases, including devices that operate with different polarizations and at single or broadband wavelengths.  Further code refactorization, such as that featured in Ref. \cite{su2019nanophotonic}, can provide added modularity and even interchangability of electromagnetic solvers.  Code refinement for diffractive optics optimization, including refractorization, will be the subject of future study. 

\paragraph{GLOnets.}  GLOnet is a global optimizer based on the training of a generative neural network and can output highly efficient topology-optimized metasurfaces that operate near or at the physical limits of structured media engineering\cite{jiang2019free, jiang2019simulator}. A key feature of the optimization approach is that the network initially generates a distribution of devices that broadly samples the design space, and it then shifts and refines this distribution towards favorable design space regions over the course of optimization. Training is performed by calculating the forward and adjoint electromagnetic simulations of outputted devices and using the subsequent efficiency gradients for backpropagation.  The GLOnets code posted to MetaNet is from Ref. \cite{jiang2019simulator}. GLOnet code consists of two parts. First, the neural network is defined and trained in Python utilizing the PyTorch library.  Second, this network interfaces with Matlab code that performs the electromagnetic simulations using the RCWA solver Reticolo. Detailed documentation of GLOnets is available online: \href{https://github.com/jiaqi-jiang/GLOnet/blob/master/README.md}{https://github.com/jiaqi-jiang/GLOnet/blob/master/README.md}.

\section{Applications}

We present three analyses of the datasets in MetaNet: we compare the performance of 2D versus 3D devices, we demonstrate how large-area aperiodic devices can be constructed by stitching together multiple devices from the MetaNet library, and we utilize principal component analysis to map out the design space and identify key geometric features of high-performance devices.  

\subsection{Benchmarking 2D and 3D devices}

\begin{figure}[H]
    \centering
    \includegraphics[width =  \linewidth]{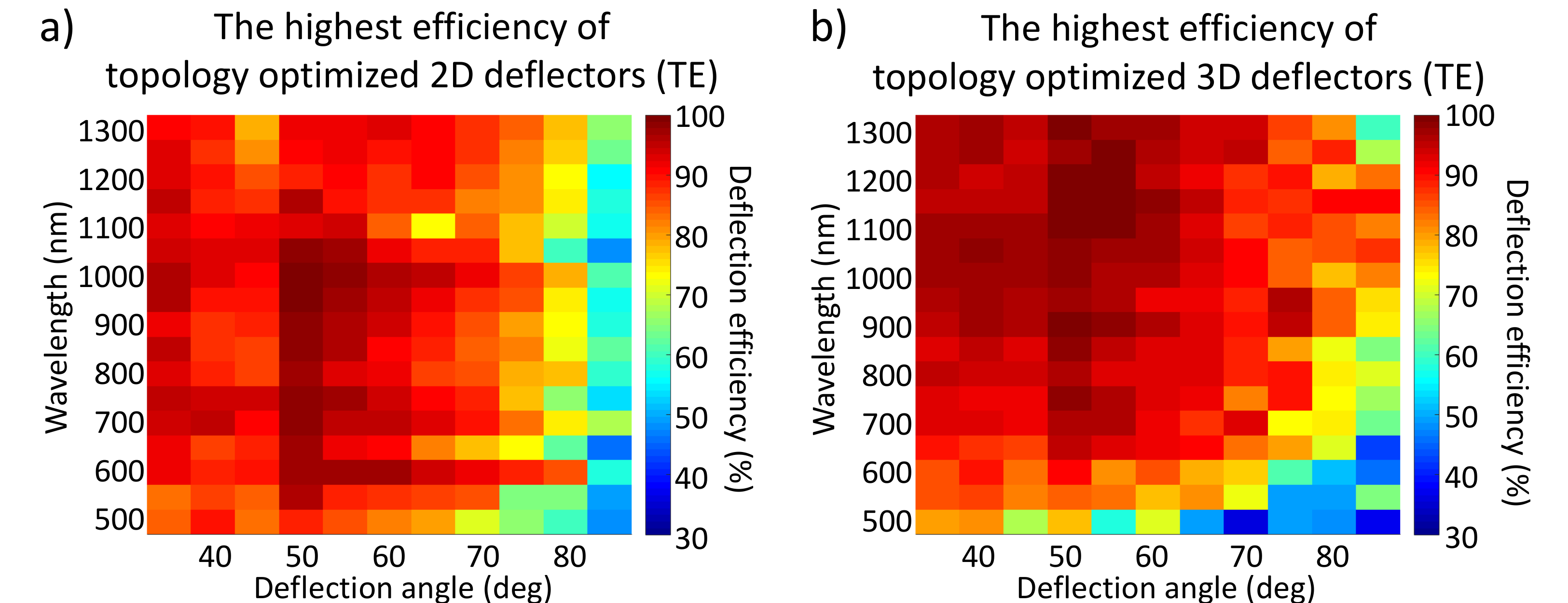}
    \caption{Comparison of 2D and 3D deflectors. The highest efficiency gradient-based topology-optimized deflectors are plotted in (a) for 2D devices and (b) for 3D devices.}
    \label{fig:benchmark}
\end{figure}

With the MetaNet library, it is straightforward to directly compare and benchmark different methods and device types.  Here, we compare 2D and 3D metagratings performing the same task of deflecting light of specific wavelength to a target angle. In Figure \ref{fig:benchmark}, we plot the efficiencies of the highest performing 2D and 3D deflectors generated by gradient-based topology optimization featured in the database. During optimization, the 2D devices are restricted to a minimum feature size of 30 nm.  As it is difficult to add strict minimal feature size constraint on 3D devices, the 3D devices are optimized with robustness control of 5 nm edge deviation \cite{wang2019robust} to ensure fabricability.  All devices have the same thickness of 325 nm and operate with TE polarization. We compare deflectors across a wavelength range of 500 to 1300 nm and deflection angles from 35 to 85 degrees.  

The average efficiency across all wavelength-angle pairs for 2D devices is 84.7\% (Figure \ref{fig:benchmark}a), compared to 86.7\% for 3D devices (Figure \ref{fig:benchmark}b). The difference in efficiency between 2D and 3D devices is generally expected given that the design space for 3D devices is much larger than that for 2D devices.  
2D optimized deflectors have higher performance at smaller deflection angles than those at larger deflection angles. The worse performance at large angles, where the unit cell is the smallest and nearly wavelength-scale, suggests a physical limitation to deflection efficiencies of 2D devices at large deflection angles.
3D deflectors tend to have good performance at longer wavelengths and, with a larger design space than that for 2D devices, can achieve relatively higher efficiencies at large deflection angles.  However, optimized 3D devices are less efficient than 2D devices at shorter wavelengths, indicating that the design space is highly complex with many local optima, making it difficult to discover high-performance devices from initially random dielectric distributions.  This simple demonstration indicates that success in identifying high performing devices through optimization is strongly dependent on the detailed optimization landscape.

\subsection{Large-area design}
While the MetaNet library currently contains only unit cells of periodic deflectors, it is possible to stitch many of these unit cells together from differing devices to create an aperiodic device such as a metalens. This concept of constructing metasurfaces through the stitching of wavelength-scale unit cells has been previously demonstrated in \cite{phan2019high} and \cite{Jiang:18}. Here, we stitch high deflection angle unit cells from MetaNet, which are designed to deflect TE polarized light to angles ranging from 55$^{\circ}$ to 75$^{\circ}$,  to create an off-axis metalens.

\begin{figure}[H]
    \centering
    \includegraphics[width =  100mm]{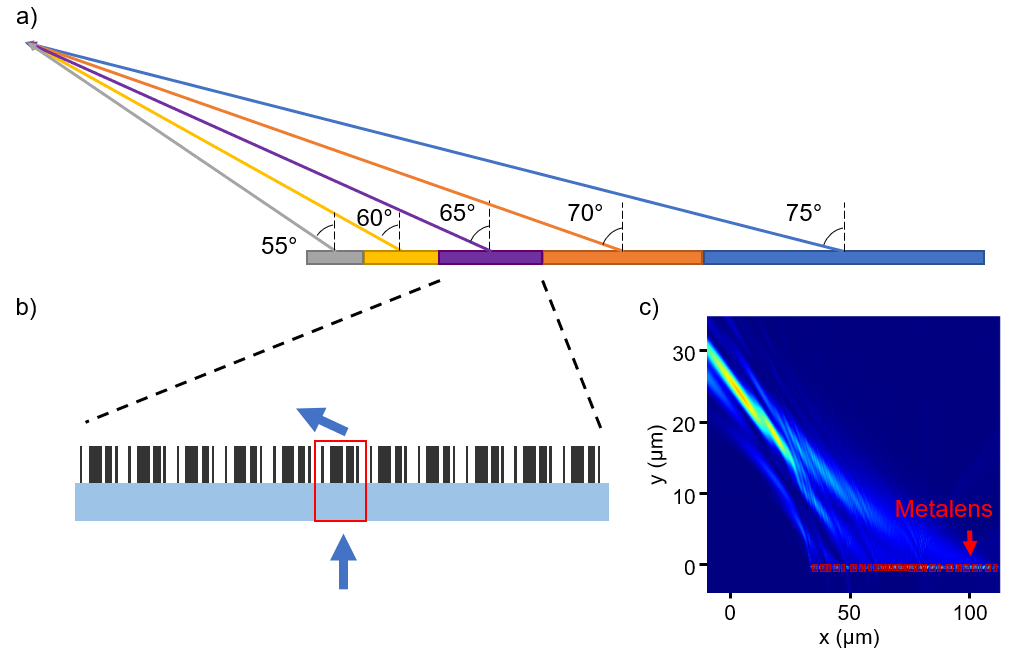}
    \caption{Off-axis metalens design.  (a) Schematic of stitched deflector elements that produce the off-axis lens. Each color represents a finite segment of a periodic deflector device.  (b) Example side view profile for the 65$^{\circ}$ deflector section of the lens. The device is taken directly from the MetaNet dataset. (c) Intensity profile of the focal point of the off-axis metalens. The device region is indicated in red.}
    \label{fig:stitching}
\end{figure}

To construct this device, we first decompose the continuous off-axis lens phase profile into segments that are linearized and represented by a single deflection angle, as shown in Figure \ref{fig:stitching}a. Each segment is then replaced by multiple periods of a corresponding 2D deflector from MetaNet, as shown in Figure \ref{fig:stitching}b.  In using a finite number of unit cells from a periodic deflector, it is important to use at least five unit cells per segment, since the individual unit cells are only designed to operate only in a periodic array.   Finally, in order to achieve constructive interference at the focal point, the phase relationship between each deflector section is adjusted by laterally shifting each section.

A simulation of intensity profile of the focal point of the completed device is shown in Figure \ref{fig:stitching}c. The focal point is clearly visible, though there are visible aberrations due to light leaking through empty regions between the deflector segments and  errors resulting from the linearization of the lens phase profile.  Improvements to device performance can be achieved by performing gradient-based optimization of the device after stitching, as well as by stitching together scattering elements specifically designed for aperiodic device construction \cite{phan2019high}.  We anticipate the inclusion of aperiodic scattering elements in future versions of MetaNet.

\subsection{Statistical analysis of dataset}

Principle component analysis (PCA) is a technique for analyzing high-dimensional datasets by reducing them to a more manageable number of dimensions.  In this section, we present the PCA of 2D deflectors generated by unconditional GLOnets. \cite{jiang2019simulator} The lower dimensional PCA representation allows us to directly visualize the variance in GLOnet-generated device patterns.  

\begin{figure}[H]
    \centering
    \includegraphics[width =  \linewidth]{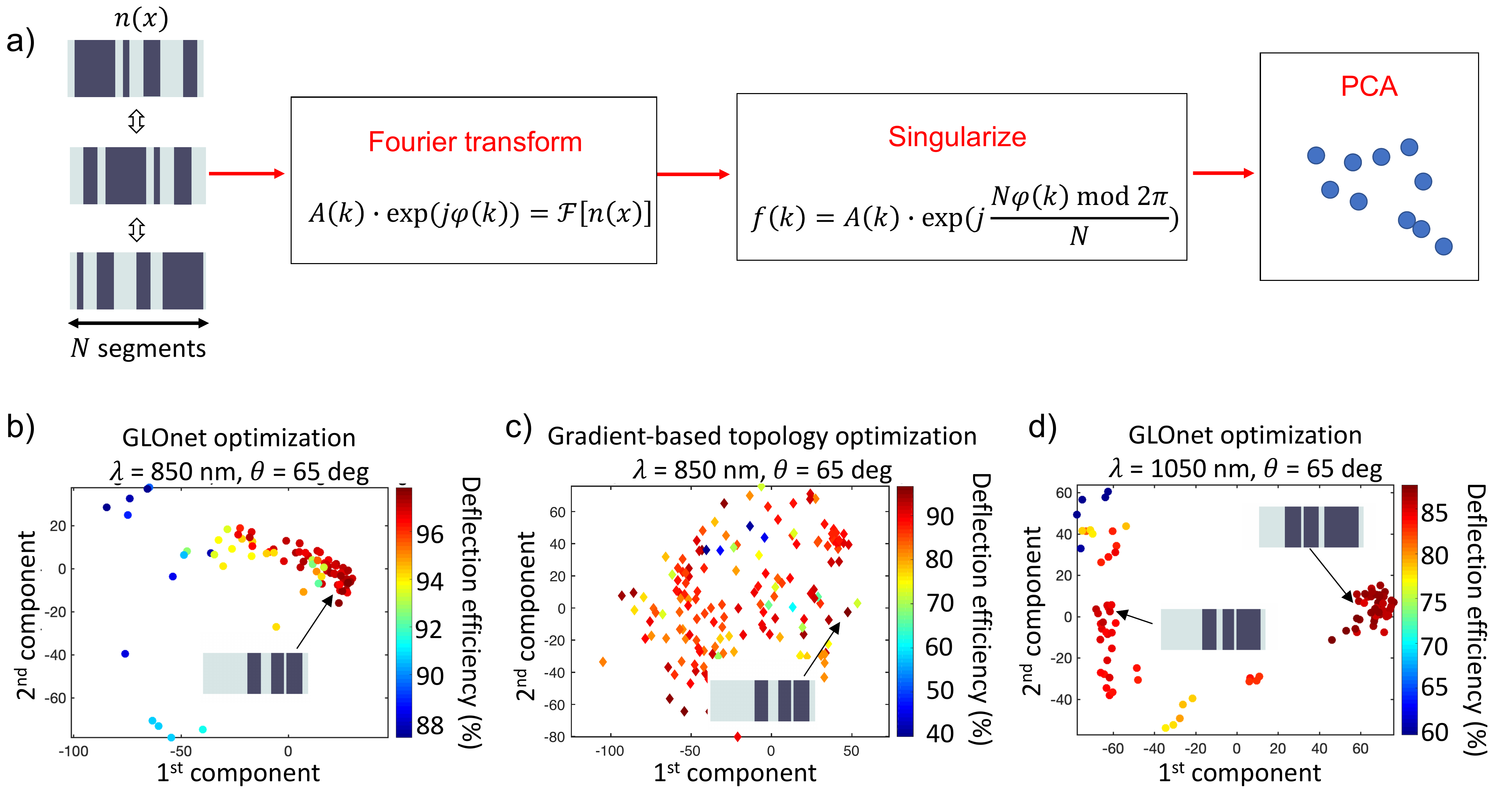}
    \caption{PCA analysis of metagratings.  (a) Preprocessing of a periodic grating pattern $n(x)$ for PCA. (b) PCA of GLOnet-optimized devices for wavelength of 850 nm and deflection angle of 65 degrees. (c) PCA of GLOnet optimized devices for wavelength of 1050 nm and deflection angle of 65 degrees. (d) PCA of gradient-based topology optimized devices for wavelength of 850 nm and deflection angle of 65 degrees.}
    \label{fig:pca}
\end{figure}

An important consideration in this analysis is that the individual metagrating unit cells are invariant to arbitrary spatial translations and wrapping.  As an example, the left part of Figure \ref{fig:pca}a shows three physically identical unit cell patterns exhibiting different levels of spatial translation and wrapping.  If we naively apply conventional PCA directly to these patterns, these equivalent patterns will appear as distinctive devices.  In order to account for the translational invariance of our system, we preprocess our devices in a manner outlined in Figure \ref{fig:pca}a.  Given a periodic pattern, we denote the refractive index value at position $x$ by a $1\times N$ vector $n(x)$. We first perform a Fourier transform on $n(x)$:
\begin{equation}
    A(k)\cdot \exp\left(j\phi(k)\right) = \mathcal{F}[n(x)]
\end{equation}
$A(k)$ and $\phi(k)$ represent the amplitude and the phase of the $k^\text{th}$ Fourier coefficient.  Spatial translations of the patterns manifest as phase differences in the Fourier domain:
\begin{equation}
    \mathcal{F}[n(x - m)] = \mathcal{F}[n(x)] \cdot  \exp{\left(-j\frac{2\pi m k}{N} \right)}
    \label{eq1}
\end{equation}
We can eliminate this phase difference and enforce translational invariance by taking the Fourier coefficients to the $N^\text{th}$ power:
\begin{equation}
    \mathcal{F}^N[n(x - m)] = A^N(k) \cdot \exp(j N\phi(k)) = \mathcal{F}^N[n(x)]
    \label{eq2}
\end{equation}
To normalize the amplitudes of these Fourier coefficients so that they accurately represent the original patterns, we take the $N^{\text{th}}$ root of Equation \ref{eq2} and constrain the phase from $[0, 2\pi/N]$, so that the Fourier coefficients are transformed as $f(k)$ :
\begin{equation}
    f(k) = A(k)\cdot \exp \left(j \frac{N \phi(k)\ \text{mod}\ 2\pi}{N} \right)
    \label{eq3}
\end{equation}
$N \phi(k)\ \text{mod}\ 2\pi$ gives the reminder after division of $N \phi(k)$ by $2\pi$.  

With our translationally invariant device representations, we apply PCA to better understand and map out the distribution of globally optimized devices within the design space. For a given operating wavelength and deflection angle pair, we first train twenty GLOnets and build a "globally optimized" dataset by generating five hundred devices and selecting the top five devices from each GLOnet. We then use the aforementioned preprocessing procedure to transform this dataset from $x(n)$ to $f(k)$.  Finally, we  perform PCA on $f(k)$. 

Representative PCA results are shown in Figure \ref{fig:pca}. Figure \ref{fig:pca}b shows results for devices operating at a wavelength of 850 nm and a deflection angle of 65 degrees.  The majority of GLOnets converge to the same global optimum (i.e., cluster of red dots in the figure), which is a device possessing three ridges that operates with an efficiency of 97\%.  This plot indicates the presence of a single global optimum and the ability of GLOnets to consistently identify the global optimum.

For comparison, we perform PCA on one hundred devices produced by gradient-based topology optimization, shown in Figure \ref{fig:pca}c.  These devices have the same design parameters as the device in Figure \ref{fig:pca}b.  The resulting distribution is markedly different with a diversity of optimized patterns spanning the entire reduced design space.  The best device is identified in the inset and is the same as the best GLOnet-generated device.  This plot visually illustrates that gradient-based topology optimization is a local optimizer and that many different local optima are identified when the optimizer is initiated with a random dielectric distribution.
While the global optimum is identified using many instances of local optimization, it is the optimal solution for only one in one hundred local optimizations for this particular problem.

We also find that some device parameter spaces support local optima with efficiencies very close to that of the global optimum.  At a wavelength of 1050 nm and deflection angle of 65 degrees, our analysis of GLOnet-based devices indicates that there exist two optima with very high efficiencies (Fig. \ref{fig:pca}d). The group of devices on the right side of the plot is clustered around the global optimum, which has an efficiency of 88\%.  The group of devices on the left side of the plot is clustered around a local optimum with an efficiency of 86\%, which is very close to that of the global optimum.  This plot suggests that for some device parameter spaces, high performance locally-optimal devices exist with efficiencies very close to that of the global optimum.  For practical design purposes, it is often sufficient to design around these local optima.  Furthermore, the plot shows that more devices cluster around the global optimum than around the local optimum, which is a good indication of the global optimum discovery capabilities of GLOnet.

\begin{figure}[H]
    \centering
    \includegraphics[width =  \linewidth]{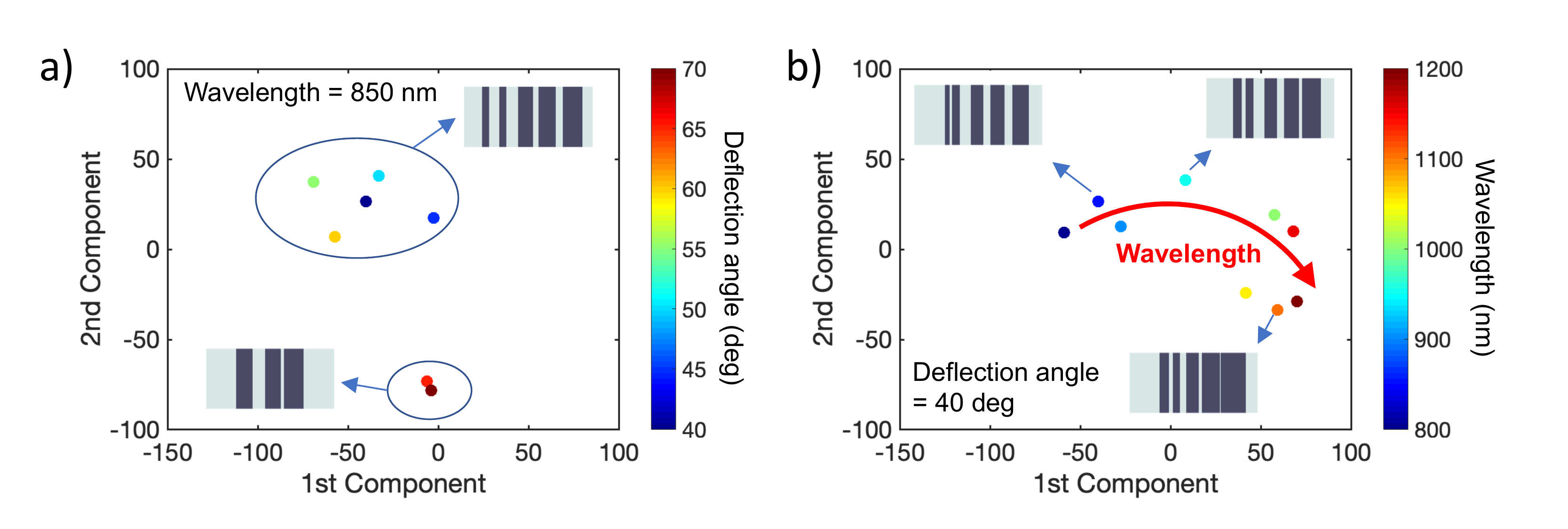}
    \caption{Device trends in the operating wavelength-deflection angle parameter space.  (a) PCA of globally-optimized devices of various deflection angles. (b) PCA of globally-optimized devices of various wavelengths.}
    \label{fig:clustering}
\end{figure}

Finally, we show that PCA can be used to analyze the evolution of globally-optimized devices in the operating wavelength-angle parameter space. In Figure \ref{fig:clustering}a, we perform PCA on globally-optimized devices with a fixed operating wavelength of 850 nm and deflection angles ranging from 40 to 70 degrees.  We find that as we increase the deflection angle, the topology of these devices jumps from one type to another, with a transition angle between 60 degrees and 65 degrees.  As such, fundamentally different light-matter interactions are required for light deflection at different angles.  On the other hand, when we perform PCA on devices with a fixed deflection angle of 40 degrees and wavelengths varying from 800 to 1200 nm (Figure \ref{fig:clustering}b), the device topology adiabatically changes as a function of wavelength.  The smooth geometric perturbations that follow as a function of wavelength are also captured in Movie S1.  These sets of devices therefore operate with related light-matter interactions and are smoothly interconnected within the device operating parameter space.  The close relationship between these globally optimal devices also indicates that they can be likely found by applying local gradient-based optimization to a single known globally optimal device operating at one wavelength-angle pair, or by a single conditional GLOnet.  These alternative optimization methods are a more computationally efficient pathway than running multiple unconditional GLOnets.  In general, an understanding of how photonics design spaces are partitioned by different regions of related optima will help streamline the device optimization process with high computational efficiency.

\section{Conclusions and Outlook}
We have introduced the MetaNet database as a tool to facilitate design in the nanophotonics community. Currently, we have uploaded over one hundred thousand designs of metagratings, created using both gradient-based topology optimization and global topology optimization, and have categorized the devices by geometry and function.  We have also demonstrated applications of MetaNet using our own metagrating dataset by performing aperiodic metasurface device design and using principal component analysis to better understand the design space. 

Beyond simply adding more devices and device categories, MetaNet can be still be improved in a variety of ways. For example, if we wish to utilize the maximum potential of MetaNet for machine learning, MetaNet should include not only final device patterns but also optimization trajectory information such as the intermediate patterns and gradients.  Ultimately, we anticipate that contributions from the broader community, together with coordinated dialogue and agreement in specifying good model systems and device parameters for study, will be required for MetaNet to have maximum efficacy in our community.

The creation of a central database for photonics devices raises questions regarding standardization and data-sharing that must be collectively addressed by the photonics community in the future.  For instance, tracking and quantifying the minimum feature size and robustness in freeform devices is important because there exist practical limits to fabrication resolution and intrinsic trade offs between feature size, robustness, and device performance.  However, there currently is no well-defined metric to specify and compare features sizes in freeform devices.  Another issue is the non-uniformity of data storage, as different research groups often use entirely different file formats. For the current dataset, we have chosen the HDF5 format, the primary storage format of Matlab's ".mat" data files in newer versions. HDF5 is also supported by Python, Java, C, Julia, and many other common languages.  This format streamlines data management by allowing device layout data to be readily combined with other device-relevant data.  However, other formats such as GDSII files are more accessible and more conducive to implementation in a fabrication write file. The trades offs between various formats will have to be discussed, and it will likely be necessary to develop new software that can quickly convert design files between different file formats under a fixed set of conventions. 

Ultimately, with MetaNet, we hope to catalog ensembles of simple photonic device designs and design codes that are representative of major application areas in photonics.  An open question, which we hope will be addressed through dialogue with the photonics community, is to identify proper model systems that capture the design challenges and design space of these major application areas.  Specific candidate systems that we envision to be appropriate for MetaNet include: silicon photonic wavelength splitters, which are a model system for chip-scale components; wavelength filters, which are a model system for thin film dielectric stacks; plasmonic scatterers, which are a model system for tunable scattering systems; and silicon grating couplers, which are a specific device essential for many technologies.

\section*{Funding}
The simulations were performed in the Sherlock computing cluster at Stanford University. This work was supported by the U.S. Air Force under Award Number FA9550-18-1-0070, the Office of Naval Research under Award Number N00014-16-1-2630, and the David and Lucile Packard Foundation.

\section*{Disclosures}
The authors declare no conflicts of interest.



\end{document}